\documentclass[aps,pra,showpacs]{revtex4}
\usepackage{graphicx}

\begin{document}

\title{Complementary classical fidelities
as an efficient criterion for the evaluation
of experimentally realized quantum operations}


\author{Holger F. Hofmann}
\email{h.hofmann@osa.org}
\affiliation{
Graduate School of Advanced Sciences of Matter, Hiroshima University,
Kagamiyama 1-3-1, Higashi Hiroshima 739-8530, Japan}

\begin{abstract}
It is shown that a good estimate of the fidelity of an experimentally
realized quantum process can be obtained by measuring the outputs
for only two complementary sets of input states. The number of measurements
required to test a quantum network operation is therefore only twice
as high as the number of measurements required to test a corresponding
classical system.
\end{abstract}


\maketitle


One of the greatest challenges in quantum information science is the
experimental realization of well-controlled operations on increasingly
complex quantum systems. In particular, quantum computation is based on
the implementation of networks of universal quantum gates operating
at low noise \cite{Nie00}. Recently, there have been several successful
experimental demonstrations of quantum controlled-NOT gates that could
serve as essential elements in future quantum computation networks
\cite{Pas03,SKa03,Lei03,Bri03,Gas04,Hua04,Zha05}. Since all of these devices
operate at non-negligible noise levels, there has also been an increasing
interest in the quantification of noise and the development of efficient
criteria for the comparison of different experiments
\cite{Poy97,Chu97,Nie03,Har03,Bri04,Gil04}.
However, the criteria presently discussed in the literature are
mostly based on theoretical considerations, and experimentalists have
usually evaluated the performance of their devices on an "{\it ad hoc}"
basis instead of applying the more complicated and often non-intuitive
procedures necessary to obtain an evaluation fulfilling the theoretical
requirements for a good measure (see \cite{Gil04} for an interesting discussion
of this problem and an overview of error measures for quantum processes).
In order to bridge this gap between the experimentalists intuition and the
theorists requirements for a good error measure, it may thus be useful to
investigate the possibility of estimating the performance of quantum
devices based on a minimal number of well-defined experimental tests.

In the following, it is shown that any unitary transform $\hat{U}_0$ is
uniquely defined by its observable effects on only two complementary
sets of orthogonal input states \cite{qcom}. The performance of any device
implementing the unitary transform $\hat{U}_0$ can therefore be tested by
measuring the classical fidelities of these two complementary operations.
The relationship between the complementary fidelities and the
overall process fidelity is discussed and upper and lower bounds for an
estimate of the process fidelity are given \cite{fidelity}.
An estimate of the process fidelity
for an $N$-level system can thus be obtained from only $2N$ measurement
probabilities, corresponding to the successful performance of two well-defined
classical operations on the respective sets of orthogonal input states.


If the desired operation of a quantum device is described by the unitary operator
$\hat{U}_0$, the expected outcomes for a specific set of orthogonal input states
$\{\mid n \rangle \}$ are given by
\begin{equation}
\label{eq:basic}
\hat{U}_0 \mid n \rangle = \mid f_n \rangle.
\end{equation}
Since $\hat{U}_0$ is unitary, the output states also form an orthogonal set
$\{\mid f_n \rangle \}$. It is therefore possible to verify the operation
described by equation (\ref{eq:basic}) by a conventional von Neumann measurement
of the output \cite{note}.
For an experimental realization of the intended unitary operation $\hat{U}_0$,
the fidelity of this classically defined operation is equal to the
average probability of obtaining the correct output for each of
the $N$ possible input states. If the actual experimental process is described
by the linear map $\hat{\rho}_{\mbox{out}} = E(\hat{\rho}_{\mbox{in}})$,
this classical fidelity is given by
\begin{eqnarray}
\label{eq:classn}
F_{n\to f_n}
&=& \frac{1}{N} \sum_{n=1}^N
\langle n \mid \hat{U}_0^\dagger
\; E(\mid n \rangle\langle n \mid)\;
\hat{U}_0 \mid n \rangle
\nonumber
\\
&=& \frac{1}{N} \sum_{n=1}^N
\langle f_n \mid  E(\mid n \rangle\langle n \mid)  \mid f_n \rangle
\nonumber
\\
&=& \frac{1}{N} \sum_{n=1}^N p(f_n | n).
\end{eqnarray}
Since the classical concept of fidelity represents a very intuitive
test of device performance, it has been commonly used to characterize
the operation of experimental quantum gates in the computational basis
\cite{SKa03,Bri03,Gas04}.
However, it is generally recognized that the classical fidelity is
not sufficient as an experimental criterion for the successful
implementation of $\hat{U}_0$ since it is not sensitive to quantum
coherence between different input and output states.
In particular, a fidelity of one can be obtained for a large number of
processes $E(\hat{\rho}_{\mbox{in}})$, some of which can actually have a
process fidelity of zero with respect to the intended unitary operation
$\hat{U}_0$.

To analyze what kind of information about the experimental process $E(\hat{\rho}_{\mbox{in}})$ is actually obtained from a measurement of the
classical fidelity defined by equation (\ref{eq:classn}), it is useful to
consider a set of $N$ orthogonal
quantum processes $U_q$ with a fidelity of $F_{n\to f_n} = 1$ \cite{ortho}.
A convenient expression for such a set of orthogonal processes can be
defined by
\begin{equation}
\label{eq:uq}
\hat{U}_q \mid n \rangle = \exp[-i \frac{2\pi}{N} q n] \mid f_n \rangle.
\end{equation}
Note that this set of orthogonal unitary transformations is not unique,
since the definition of phase for the output states $\mid f_n \rangle$
is quite arbitrary. In this sense, equation (\ref{eq:uq}) only gives
an example of how to construct an orthogonal set of $N$ unitary
transformations with a classical fidelity of one for the operation $n\to f_n$.
The experimental process $E(\hat{\rho}_{\mbox{in}})$ can then be expanded
in terms of a complete set of $N^2$ orthogonal basis operators $\{\hat{U}_q\}$,
where the first $N$ basis operators are defined according to equation
(\ref{eq:uq}), and the remaining $N(N-1)$ operators can be any set of
orthogonal unitary operators spanning the remaining process space,
\begin{equation}
\label{eq:expand}
E(\hat{\rho}_{\mbox{in}}) = \sum_{q,r=0}^{N^2-1} \chi_{q,r}
\hat{U}_q \hat{\rho}_{\mbox{in}}\hat{U}_r^\dagger.
\end{equation}
The fundamental properties of this expansion are most easily understood by
considering the application of $E(\hat{\rho}_{\mbox{in}})$ to a maximally
entangled state $\mid \phi \rangle_{\mbox{\small A,R}}$ of the system A
and a reference R. If $E$ is applied only to A (that is, $E \otimes I$ is
applied to the joint system of A and R), the process matrix is then
equal to the density matrix of the output state for the orthogonal basis
states $\{ \hat{U}_q \otimes \hat{1} \mid \phi \rangle_{\mbox{\small A,R}}\}$
generated by applying the basis operators $\{\hat{U}_q\}\otimes\hat{1}$
to the pure state input $\mid \phi \rangle_{\mbox{\small A,R}}$.
From this observation, it follows that the process matrix is a positive hermitian matrix with a
trace of one (or less for conditional operations with a limited probability
of success). Moreover, the process fidelity can be defined as the
probability of obtaining the output state $\hat{U}_0 \otimes \hat{1}
\mid \phi \rangle_{\mbox{\small A,R}}$ corresponding to the application of the
ideal process $\hat{U}_0$ to system A of the pure state input
$\mid \phi \rangle_{\mbox{\small A,R}}$. Since this
probability is equal to the corresponding diagonal element of the
process matrix, the overall process fidelity is then given by
$F_{\mbox{\small process}}=\chi_{0,0}$.

Using the expansion given by equation (\ref{eq:expand}), the classical
fidelity $F_{n\to f_n}$ can now be related directly to the
elements $\chi_{q,r}$ of the process matrix,
\begin{equation}
\label{eq:nfn}
F_{n\to f_n} = \chi_{0,0} + \sum_{q=1}^{N-1} \chi_{q,q}.
\end{equation}
In terms of the linear algebra of process expansions, the classical
fidelity $F_{n\to f_n}$ corresponds to a projective measure
of the process components that lie within the $N$-dimensional subspace
of the $N^2$-dimensional process-space spanned by the orthogonal basis
$\{\hat{U}_q\}$.
Since this subspace is larger than the one dimensional subspace
representing the ideal operation,
the classical fidelity $F_{n\to f_n}$ is always equal to or greater
than the process fidelity given by $\chi_{0,0}$. Each classical
fidelity thus provides an upper bound for the overall process
fidelity \cite{Bar98}.

In order to experimentally distinguish the $N$ operations $\hat{U}_q$
with classical fidelities of $F_{n\to f_n}=1$ from each other, it is
necessary to change the input basis. Optimal distinguishability is
achieved when the output states of different $\hat{U}_q$ for the same
input state are orthogonal to each other. This condition can be fulfilled
by complementary sets of input states $\mid k^\prime \rangle$
with $|\langle n \mid
k^\prime \rangle|^2=1/N$ for all $n$ and $k$, as e.g. given by
\begin{equation}
\mid k^\prime \rangle =
\frac{1}{\sqrt{N}}\sum_{n=1}^N \exp[-i \frac{2\pi}{N} k n] \mid n \rangle.
\end{equation}
For this set of states, the unitary operation $\hat{U}_0$ defines a
second classical function, given by
\begin{equation}
\label{eq:comp}
\hat{U}_0 \mid k^\prime \rangle = \mid g_k^\prime \rangle,
\end{equation}
where the output states $\mid g_k^\prime \rangle$ are also
complementary to the output states $\mid f_n \rangle$ according to
\begin{equation}
\mid g_k^\prime \rangle = \frac{1}{\sqrt{N}} \sum_{n=1}^N \exp[-i \frac{2\pi}{N} k n] \mid f_n \rangle.
\end{equation}
Since these output states are maximally sensitive to the quantum phases
between the components $\mid f_n \rangle$, the effects of different unitary
operations $\hat{U}_{q < N}$ on the quantum phases of $\mid f_n \rangle$ becomes
directly observable in the output basis $\mid g_k^\prime \rangle$.
Specifically,
\begin{equation}
\label{eq:compshift}
\hat{U}_{q < N} \mid k^\prime \rangle = \mid g_{k+q}^\prime \rangle.
\end{equation}
Thus the output states for different operations $\hat{U}_q$ are indeed
orthogonal, making the operation on the complementary input states
$\mid k^\prime \rangle$ given by equation (\ref{eq:comp}) ideal for the
task of distinguishing the operations $\hat{U}_{0 < q < N}$
with $F_{n\to f_n}=1$ from the intended operation $\hat{U}_0$.

The classical fidelity of the complementary operation $k \to g_k$
can be obtained experimentally by
\begin{eqnarray}
\label{eq:classk}
F_{k\to g_k} &=&
\frac{1}{N} \sum_{k=1}^N
\langle k^\prime \mid \hat{U}_0^\dagger
\; E(\mid k^\prime \rangle\langle k^\prime \mid)\;
\hat{U}_0 \mid k^\prime \rangle
\nonumber
\\ &=&
\frac{1}{N} \sum_{k=1}^N
\langle g_k^\prime \mid E(\mid k^\prime \rangle\langle k^\prime \mid)
\mid g_k^\prime \rangle
\nonumber
\\
&=& \frac{1}{N} \sum_{k=1}^N p(g_k | k).
\end{eqnarray}
Again, it is possible to find $N$ orthogonal operations that all have
$F_{k\to g_k}=1$. However, the only operation that has both $F_{n \to f_n}=1$
and $F_{k\to g_k}=1$ is $\hat{U}_0$, since
\begin{equation}
\langle g_k^\prime \mid \hat{U}_{0 < q < N} \mid k^\prime \rangle =
\langle g_k^\prime \mid g_{k+q}^\prime \rangle = 0.
\end{equation}
This relation also implies that any unitary operation
$\hat{U}_{k\to g_k}$ with a fidelity of $F_{k\to g_k}=1$ is
orthogonal to the operations $\hat{U}_{0 < q < N}$, since
\begin{eqnarray}
\mbox{Tr}\{\hat{U}_{k\to g_k}^\dagger \hat{U}_{0 < q < N}\}
&=& \sum_k |\langle k^\prime \mid \hat{U}_{k\to g_k}^\dagger
\hat{U}_{0 < q < N} \mid k^\prime \rangle|^2
\nonumber \\
&=& \sum_k |\langle g_k^\prime \mid g_{k+q}^\prime \rangle|^2 = 0.
\end{eqnarray}
It is therefore possible to identifying the remaining $N-1$
orthogonal operations having classical fidelities of $F_{k\to g_k}=1$
with the basis operators $\hat{U}_{N}$ to $\hat{U}_{2(N-1)}$.
In fact, it is possible to explicitly construct an orthogonal
set of unitary operators in close analogy with equation
(\ref{eq:uq}),
\begin{equation}
\label{eq:moreuq}
\hat{U}_{N \leq q \leq 2(N-1)} \mid k^\prime \rangle =
\exp[-i \frac{2\pi}{N} (q+1) k] \mid g_k^\prime \rangle.
\end{equation}
The complementary classical fidelity
$F_{k\to g_k}$ can then be expressed in terms of the process matrix elements
$\chi_{q,r}$ of equation ($\ref{eq:expand}$) as
\begin{equation}
\label{eq:kgk}
F_{k\to g_k} = \chi_{0,0} + \sum_{q=N}^{2(N-1)} \chi_{q,q}.
\end{equation}
In terms of the linear algebra of process expansions, the complementary
fidelity $F_{k\to g_k}$ thus evaluates the component of the process
in an $N$-dimensional subspace of the $N^2$ dimensional process space
that only overlaps with the subspace defined by $F_{n\to f_n}$ in the
ideal process $\hat{U}_0$. Therefore, the maximal total fidelity
$F_{n \to f_n} + F_{k \to g_k}$ cannot exceed one unless there is a
non-vanishing contribution from the ideal process $\hat{U}_0$.

Based on these results, it is possible to derive an estimate of the
process fidelity $F_{\mbox{\small process}}=\chi_{0,0}$ from the
measured results for the classical fidelities $F_{n \to f_n}$ and
$F_{k \to g_k}$. Since the process fidelity is by definition equal
to the process matrix element $\chi_{0,0}$, the relationship between
the classical fidelities and the process fidelity is given by equations
(\ref{eq:nfn}) and (\ref{eq:kgk}). These equations show that the
classical fidelities can each be interpreted as sums of process fidelities
for $N$ orthogonal (and therefore distinguishable) processes.
If the two complementary classical fidelities are added, only the
intended process $\hat{U}_0$ contributes twice. The lower bound of the
process fidelity is therefore equal to the amount by which the
total fidelity $F_{n \to f_n} + F_{k \to g_k}$ exceeds one,
\begin{equation}
\label{eq:Fmin}
F_{n \to f_n} + F_{k \to g_k} - 1 \; \leq \; F_{\mbox{\small process}}.
\end{equation}
An upper bound for the process fidelity can be derived from the
minimum of the two classical fidelities, since the sum of $N$
process fidelities is necessarily equal to or greater than each
individual fidelity \cite{Sch96}. The upper bound thus reads
\begin{equation}
\label{eq:Fmax}
F_{\mbox{\small process}} \; \leq \;
\mbox{Min}\{F_{n \to f_n}, F_{k \to g_k}\}.
\end{equation}
Note that the difference between the lower and the upper bound depends
on the closeness of the maximal classical fidelity to one.
Specifically, if
$F_{n \to f_n}=1-\epsilon$ is greater than $F_{k \to g_k}$ and close
to one, the process fidelity will be found in an interval of width $\epsilon$
below the lower classical fidelity $F_{k \to g_k}$ given by
\begin{equation}
\label{eq:epsilon}
F_{k \to g_k}-\epsilon \; \leq \; F_{\mbox{\small process}} \; \leq \;
F_{k \to g_k}.
\end{equation}
The complementary classical fidelities are therefore particularly well
suited for an estimate of process fidelity if the performance in one
basis (e.g. the computational basis) is highly reliable and the main
error source is dephasing between these basis states \cite{Chr05}.

To place the results into a wider context, it may also be useful
to convert the process fidelity into the average quantum state
fidelity $\bar{F}$, as given by $\bar{F}=(N F_{\mbox{process}} +1)/(N+1)$
\cite{Hua04,Bri04,Gil04,Hor99}. The inequalities (\ref{eq:Fmin})
and (\ref{eq:Fmax}) then establish a relation between the
classical fidelities $F_{n \to f_n}$ and $F_{k \to g_k}$
obtained by averaging over a very specific limited selection of
input states, and the fidelity $\bar{F}$ obtained by averaging
over all possible pure state inputs.
It might be interesting to consider the implications of this
result for the relations between non-complementary classical
fidelities.

To illustrate the practical application of complementary classical fidelities,
it may be helpful to consider the specific example of a quantum controlled-NOT
gate. The effects of this gate on the computational basis (indicated by the
index $Z$ in the following) and an appropriate complementary basis (indicated
by the index $X$ in the following) can be given by
\begin{equation}
\begin{array}{lcr}
\hat{U}_{\mbox{\small CNOT}}\mid 0_Z;0_Z \rangle  &=& \mid 0_Z;0_Z \rangle
\\
\hat{U}_{\mbox{\small CNOT}}\mid 0_Z;1_Z \rangle  &=& \mid 0_Z;1_Z \rangle
\\
\hat{U}_{\mbox{\small CNOT}}\mid 1_Z;0_Z \rangle  &=& \mid 1_Z;1_Z \rangle
\\
\hat{U}_{\mbox{\small CNOT}}\mid 1_Z;1_Z \rangle  &=& \mid 1_Z;0_Z \rangle
\end{array}
\hspace{1.5cm}
\begin{array}{lcr}
\hat{U}_{\mbox{\small CNOT}}\mid 0_X;0_X \rangle  &=& \mid 0_X;0_X \rangle
\\
\hat{U}_{\mbox{\small CNOT}}\mid 0_X;1_X \rangle  &=& \mid 1_X;1_X \rangle
\\
\hat{U}_{\mbox{\small CNOT}}\mid 1_X;0_X \rangle  &=& \mid 1_X;0_X \rangle
\\
\hat{U}_{\mbox{\small CNOT}}\mid 1_X;1_X \rangle  &=& \mid 0_X;1_X \rangle,
\end{array}
\end{equation}
where the basis transformation corresponds to the application of a Hadamard
transformation to each qubit,
\begin{eqnarray}
\mid 0_X \rangle &=& \frac{1}{\sqrt{2}}
\left( \mid 0_Z \rangle + \mid 1_Z \rangle\right)
\nonumber \\
\mid 1_X \rangle &=& \frac{1}{\sqrt{2}}
\left( \mid 0_Z \rangle - \mid 1_Z \rangle\right).
\end{eqnarray}
The complementary classical fidelities of the quantum controlled-NOT gate thus
correspond to the fidelities of two classical controlled-NOT operations, where
the Hadamard transform of the input and output basis causes an exchange of
the roles of control and target qubit \cite{Hof04}. The complementary classical
fidelities of the quantum controlled-NOT gate can then be obtained from
eight measurement probabilities,
\begin{eqnarray}
\label{eq:fidelities}
F_Z &=& \frac{1}{4} \left(P_{ZZ|ZZ}(00|00) + P_{ZZ|ZZ}(01|01)
+P_{ZZ|ZZ}(11|10) + P_{ZZ|ZZ}(10|11) \right),
\nonumber
\\[0.2cm]
F_X &=& \frac{1}{4} \left(P_{XX|XX}(00|00) + P_{XX|XX}(11|01)
+P_{XX|XX}(10|10) + P_{XX|XX}(01|11) \right).
\end{eqnarray}
As discussed above, these eight measurement results are already sufficient
to obtain reliable estimates of the process fidelity $F_{\mbox{process}}$.
In particular, the lower bound of the process fidelity given by
$F_{\mbox{process}} \geq F_Z+F_X-1$ can be used to obtain estimates of the
gate performance for other sets of orthogonal input states, since the
classical fidelities of such operations are always greater than or equal
to the process fidelity. For example, an estimate of the entanglement
capability can be obtained by considering the classical fidelity
$F_{\mbox{\small entangle}}$ for the generation of maximally entangled
outputs if the control qubit input is an eigenstate of $X$ and the target qubit is an eigenstate of $Z$. The classical fidelity $F_{\mbox{\small entangle}}$ of this entanglement
generation process represents the average overlap of the output
states with the corresponding maximally entangled states. This average
overlap therefore defines a minimal amount of entanglement that can be
generated by the operation. In terms of the concurrence $C$, this lower
bound of the entanglement capability is given by
\begin{equation}
C \; \geq \; 2 F_{\mbox{\small entangle}} - 1.
\end{equation}
Since $F_{\mbox{\small entangle}} \leq F_{\mbox{\small process}}$, the
lower bound of the process fidelity given by $F_Z + F_X -1$ also applies
to $F_{\mbox{\small entangle}}$ and the entanglement capability can
be estimated directly by
\begin{eqnarray}
C  \; \geq \; 2 (F_Z + F_X) - 3.
\end{eqnarray}
If $F_Z=1-\epsilon$ is close to one, the gate is thus capable of generating
entanglement if $F_X$ is greater than $0.5-\epsilon$.
Note that this estimate of the entanglement capability can be obtained
without actually generating any entanglement when the device is
tested. The possibility of entanglement generation is simply a necessary
consequence of the high fidelity observed in the complementary local
operations of the quantum gate.

In summary, it has been shown that an efficient test of experimentally
realized quantum operations can be performed by measuring the
classical fidelities for only two complementary sets of orthogonal
input states. This simplified test criterion can provide good estimates
of the process fidelity and other characteristic properties of the noisy
experimental process from only $2N$ measurement probabilities.
In the case of a quantum controlled-NOT
operation, the complementary classical fidelities can be determined from
the measurement probabilities of eight pairs of local inputs and outputs.
For comparison, the precise determination of process fidelity from local
inputs and outputs reported in \cite{Bri04} was based on 71 measurement
probabilities out of the 256 probabilities required for complete quantum
process tomography.
The complementary classical fidelities therefore provide a compact and
intuitive measure of how well a given experimental device performs a
desired quantum process.

Part of this work has been supported by the JST-CREST project on
quantum information processing.


\begin{thebibliography}{xyz00}
\bibitem{qcom}
As a simple example, it may be useful to consider a quantum channel,
ideally represented by the identity operation $\hat{U}_0=\hat{1}$.
The results reported in this paper thus also apply to
fidelities in quantum communication (see [19-22]).
\bibitem{note}
In fact, equation (\ref{eq:basic}) can be interpreted in terms of
the Heisenberg picture as a transformation of the physical property
$\hat{n}$ with the eigenstates $\mid n \rangle$ into a different physical
property $\hat{f}$ with eigenstates $\mid f_n \rangle$. It therefore provides
a clear physical picture of the quantum process $\hat{U}_0$.
\bibitem{ortho}
Here, orthogonality is defined by the product trace of two operators,
that is, $\hat{A}$ and $\hat{B}$ are orthogonal if
$\mbox{Tr}\{\hat{A}^\dagger \hat{B}\}=0$.
\end{thebibliography}
\end{document}